\documentclass[12pt,preprint]{aastex}

\newcommand{\CII}{[\ion{C}{2}]~}
\newcommand{\OI}{[\ion{O}{1}]~}
\newcommand{\HII}{\ion{H}{2}~}
\newcommand{\HI}{\ion{H}{1}~}
\newcommand{\CI}{[\ion{C}{1}]~}

\begin{document}

\title{ISO-LWS observations of C$^+$ and O$^0$ lines in absorption
toward Sgr~B2\footnote{Based on observations with ISO, an ESA project with instruments
funded by ESA Member States (especially the PI countries: France, Germany,
the Netherlands and the United Kingdom) with the participation of ISAS and
NASA.}}
 
\author{C. Vastel\altaffilmark{1}}
\affil{Downs Laboratory of Physics, 320-47, California Institute of Technology, 
Pasadena, CA 91125, USA}
\author{E.T. Polehampton\altaffilmark{2}}
\affil{Department of Astrophysics, University of Oxford, Keble Road, Oxford, OX13RH, UK}
\author{J.-P. Baluteau}
\affil{Laboratoire d'Astrophysique de Marseille, CNRS \& Universit\'e de Provence, 
BP 8, F-13004 Marseille, France}
\author{B.M. Swinyard}
\affil{Rutherford Appleton Laboratory, Chilton, Didcot, Oxon, UK}
\author{E. Caux}
\affil{CESR CNRS-UPS, BP 4346, F-31028 - Toulouse Cedex 04, France}
\author{P. Cox}
\affil{Institut d'Astrophysique Spatiale, Universit\'e de Paris-Sud, F-91405 Orsay Cedex, France}
\email{Charlotte Vastel: vastel@submm.caltech.edu}
\altaffiltext{1}{also at the CESR CNRS-UPS, BP 4346, F-31028 - Toulouse Cedex 04, France}
\altaffiltext{2}{also at the Rutherford Appleton Laboratory, Chilton,
Didcot, Oxon, UK}

\date{Received {\today} /Accepted {\today}}


\begin{abstract}
High spectral resolution Fabry-P\'erot observations of the \OI 
63.2 and 145.5~$\mu$m and \CII 157.7 $\mu$m fine
structure lines are presented 
for the center of the Sagittarius B2 complex (Sgr~B2). The data were obtained 
with the Long Wavelength Spectrometer on board the Infrared 
Space Observatory (ISO). Both the \OI 63.2~$\mu$m and the 
\CII 157.7~$\mu$m lines are detected in absorption. 
The upper state level of atomic oxygen at 145.5~$\mu$m is in emission.
Whereas the \OI 63.2~$\mu$m line is seen in absorption 
over the entire wavelength range $-200$ to 100~km~s$^{-1}$, 
the \CII 157.7~$\mu$m line displays a more complex profile:
absorption occurs at velocities $<$ 20~km~s$^{-1}$ and emission
comes from the Sgr~B2 complex at velocities greater than
20~km~s$^{-1}$. Using observations of the CO isotopes and of the \HI
lines, absorption components can be
associated with many clouds along the Sagittarius B2 line of sight. 
From these data, we were able to disentangle three different layers 
which contain atomic oxygen. These layers, as predicted by PDR models, are
characterized by different forms of carbon in the gas phase, i.e. the
C$^+$ external layer, the C$^+$ to C$^0$ transition and the CO
internal layer. We derive lower limits for the column densities
of atomic carbon and oxygen of the order of $\sim$ 10$^{18}$~cm$^{-2}$ 
and 3 $\times$ 10$^{19}$~cm$^{-2}$, respectively. 
An O$^0$/CO ratio of around 2.5 is computed in the internal cores of
the clouds lying along the line of sight, which means that $\sim$ 70\%
of gaseous oxygen is in the atomic form and not locked into CO. 
The fact that the \CII 157.7 $\mu$m line is detected in absorption 
implies that the main cooling line of the interstellar medium can 
be optically thick especially in the direction of large star-forming 
complexes or in the nuclei of galaxies. This could partially account 
for the  deficiency in the \CII 157.7 $\mu$m line which has been recently found 
toward infrared bright galaxies in ISO data.
\end{abstract}


\section{Introduction}

Observations have suggested that, in some molecular clouds, most of the 
gas-phase oxygen might be in atomic form, in contradiction with predictions of 
steady state chemical models (e.g., Lee, Bettens \& Herbst 1996) which yield 
CO, O, and O$_2$ as the major oxygen bearing species in molecular clouds. First 
suggestions of a high atomic oxygen abundance were made by Jacq et al. (1990) 
and Schulz et al. (1991) in order to interpret their HDO observations of hot 
cores and quiescent clouds. These results are in accord with models which take into 
account cosmic-ray induced photo-dissociation (e.g., Jacq et al. 1990; Wannier 
et al. 1991). Further evidence has been obtained from observations of the 
\OI 63.2~$\mu$m fine structure line which has been detected in absorption 
against the far-infrared continuum of bright galactic sources, namely:
DR~21 (Poglitsch et al. 1996), Sgr~B2 (Baluteau et al. 1997) and NGC~6334V
(Kraemer et al. 1998). These observations indicate that the absorbing material
is predominantly in foreground cloud(s) where the abundance of the atomic oxygen 
is within a factor of a few of the cosmic abundance. 

Recent studies have strengthened the above results. Based on ISO-LWS Fabry-P\'erot 
data, Vastel et al. (2000) have modelled the \OI 63.2~$\mu$m line absorption 
components in the direction of the compact \HII region W49~N. Combining these 
observations with molecular (CO and its isotopes) and \HI observations, they showed that 
both molecular and atomic clouds absorb the strong continuum at 63~$\mu$m, and 
disentangled the absorption due to the molecular clouds from the absorption due 
to the atomic (\HI) clouds. They concluded that the major part of the
63.2~$\mu$m \OI
absorption is due to the cold molecular clouds along the line of sight and, through 
the computation of O$^0$/CO ratio, that in these clouds the gaseous oxygen is almost 
totally in atomic form. 

Similar results were obtained toward Sgr~B2 by Lis et al. (2001). They
found three \OI 63.2~$\mu$m absorption components corresponding to foreground clouds,
for which the oxygen content of the atomic halo gas could be estimated based 
on \HI observations. Lis et al. (2001) found that the remaining O$^0$ column 
density is correlated with the observed $^{13}$CO column density, corresponding 
to an average O$^0$/CO ratio of about 9 and to an atomic oxygen 
abundance of 2.7 $\times$ 10$^{-4}$ in the dense gas phase. 

The full Fabry-P\'erot ISO-LWS spectrum which was obtained on Sgr~B2
as part of the ISO Central Program (see Baluteau et al. {\it in preparation}, for
a detailed description) enables us to study, at high spectral resolution,
the atomic fine-structure of oxygen and carbon in this source and to 
analyze in a consistent way the profiles of the \OI lines at 63.2 
and 145.5~$\mu$m and of the \CII 157.7~$\mu$m line.  
Sgr~B2 is the most massive star-forming region of an ensemble of dense cloud 
cores in the central ($\approx \, 500$~pc) region of the Galaxy. Its estimated
mass is larger than $\rm 5 \times 10^6 \, M_{\odot}$ (Lis, Carlstrom
\& Keene 1991), and the high opacities found toward Srg~B2 
makes it one of the best candidate for absorption studies. Sgr~B2 is located at about 
8.5~kpc from the Sun (adopting the IAU distance) and has a projected distance of 
$\rm \sim 100 \, pc$ from the Galactic Center. Ground-based spectroscopic observations 
of Sgr~B2 have shown a very complex pattern of absorption features, with numerous 
components of foreground gas associated with clouds along the line of sight, covering 
a wide range of velocities (see Sect.~3.3). 

In this paper we present the observational data of the three main far-infrared 
cooling lines and of the isotopic CO lines in Sect.~2. The absorption lines arising 
from foreground clouds, which have no physical connection with Sgr~B2, 
are modeled in Sect.~3, where we try to disentangle the absorption due 
to the molecular cores from that due to the external layers of these clouds. 
We discuss these results and summarize them in Sect.~4 and 5 respectively.

\section{Observations and Results}

\subsection{ISO observations of the C$^+$ and O$^0$ lines}

Sgr~B2 has been observed with the Long Wavelength Spectrometer
(hereafter LWS; Clegg et al. 1996) on board the Infrared Space
Observatory (hereafter ISO; Kessler et al. 1996), using the high resolution mode 
(AOT L03), within the guaranteed time program ISM\_V. The whole LWS 
spectral range, from 47 to 196 $\mu$m, was covered with 36 separate 
observations (Baluteau et al. {\it in preparation}). Each observation was carried 
out using a sampling interval of a quarter of a spectral resolution 
element with each data point repeated at least three times. 

The LWS beam of approximately 80$^{''}$ was centered at 
$\alpha=17^{\rm{h}}47^{\rm{m}}21.75^{\rm{s}}$, 
$\delta=-28^{\circ}23^{'}14.1^{''}$ (J2000). At this position 
the beam encompasses the Sgr~B2 (M) source but not 
the Sgr~B2 North and South sources (following the nomenclature
of Goldsmith et al. 1992). The peak of the far-infrared emission 
is found to be centered near the Sgr~B2 (M), in the north-west direction, 
at a distance of 7$\arcsec$ and 14$\arcsec$ at 50~$\mu$m and 100~$\mu$m, respectively 
(Goldsmith et al. 1992). No significant difference in the far-infrared 
continuum emission is found when our observations are compared with other 
AOT L04 observations centered on Sgr~B2 (M). We therefore consider hereafter 
that the major part of the far-infrared emission is included within the 
LWS field of view.

\begin{figure}
\plotone{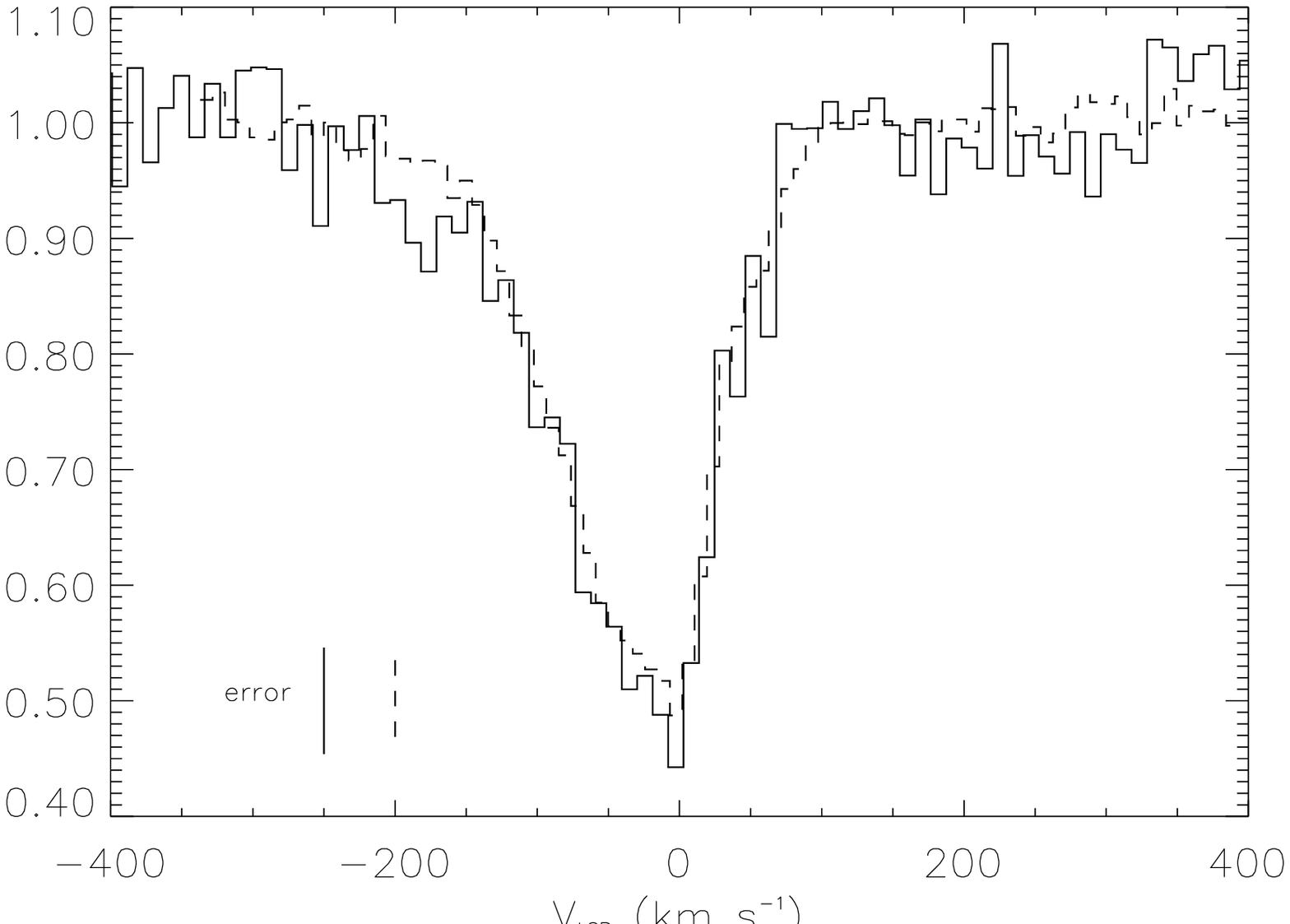} 
\caption{\OI fine-structure line at 63.2~$\mu$m toward Sgr~B2 as
observed with the ISO-LWS Fabry-P\'erot.
The L03 prime (plain line) and L04 prime data (dashed line) are compared
with an estimate of the combined systematic and random error at line
center indicated with a vertical bar. The L03 non-prime data (Fig. 2)
has a smaller error and is used in the following analysis. The y-axis
represents the line to continuum ratio. \label{f1}}
\end{figure}

\placefigure{f1}

During each observation, in the L03 operating mode, one detector was selected 
as the `prime' detector as its band pass filter included the wavelength range 
of interest. However, every detector still received some radiation if the 
combination of FP and grating settings were right. These detectors are known 
as `non-prime' and often contain useful data that can complement the prime 
data. 

The atomic oxygen fine structure line at 63.2~$\mu$m was included in the prime 
observation carried out in ISO revolution 504 on 1997 April 3. The resolving 
power of the LWS short wavelength FP (FPS) at this wavelength was determined 
by interpolating between measurements made on the ground and in orbit and found 
to be R $\sim$ 6900. The \OI 63.2~$\mu$m line was also observed in non-prime 
observations between 1997 April 3 and April 5. Three observations covered 
the line entirely and two covered the edges and adjacent continuum. All 
these observations were performed using the LWS long wavelength FP (FPL) 
outside of its nominal wavelength range. FPL had a slightly lower resolving 
power at 63~$\mu$m (R $\sim$ 5800) than FPS but achieved much better signal 
to noise ratio due to its better overall transmission. The prime and
non-prime observations were calibrated and reduced using the same
procedure as described in Polehampton et al. (2002). 
 
Figure~\ref{f1} presents the resulting spectrum at 63 $\mu$m using
L03 prime observations compared to the spectrum obtained by combining
three prime observations in L04 mode during ISO revolutions 326 and
464. In the following, we will use the spectrum obtained in 
the L03 non-prime mode, whose systematic and random noise at line
center is 2.5 times lower than in the L03 prime mode to study the 
\OI 63.2 $\mu$m line. 

\begin{figure}
\plotone{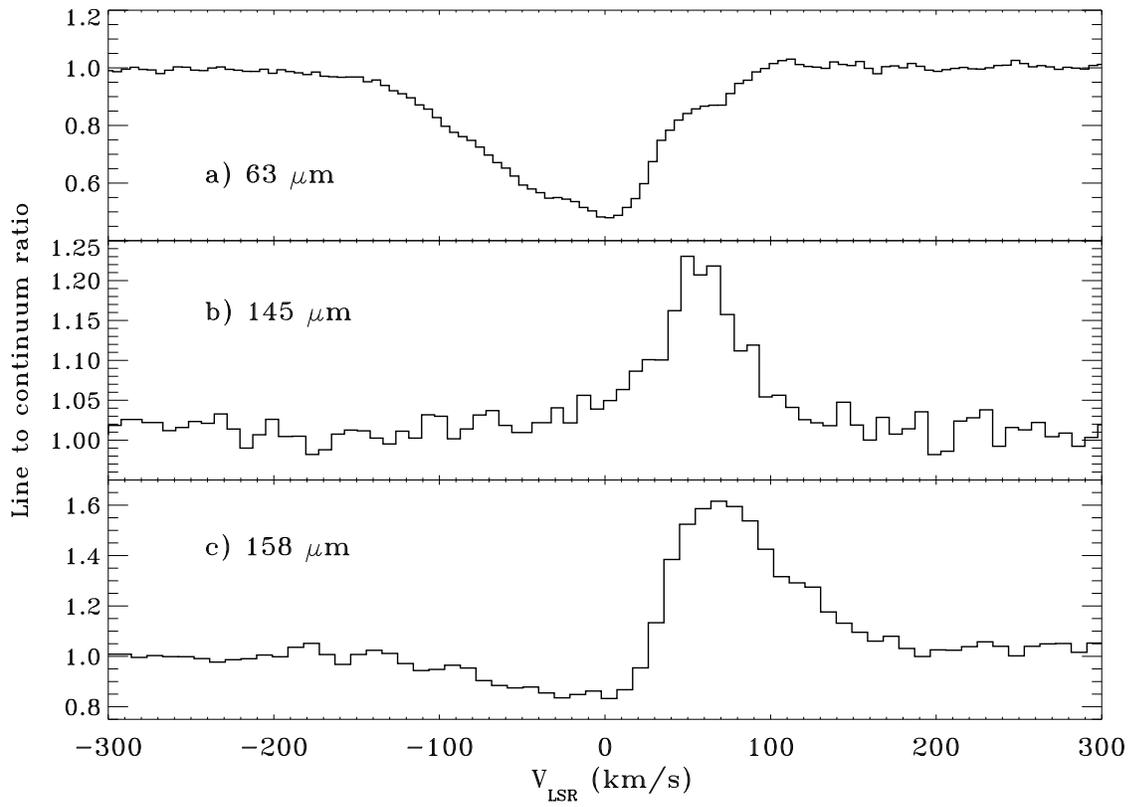}
\caption{ISO-LWS Fabry-P\'erot spectra of the \OI 63.2 $\mu$m (using
non-prime data) and 145.5~$\rm \mu m$
and \CII 157.7~$\rm \mu m$ fine structure lines toward Sgr~B2~(M).\label{f2}}
\end{figure}

\placefigure{f2}

The \OI line at 145.5~$\mu$m and the \CII line at 157.7~$\mu$m were observed 
during ISO revolutions 476 (1997 March 6) and 507 (1997 April 6), respectively, 
as prime observations. There were no good observations of either of these 
lines on non-prime detectors. These data were reduced in the same way as 
the 63.2 $\mu$m line prime observations. The resolution element 
of FPL from ground based measurement is $\sim$ 36 km s$^{-1}$ and $\sim$ 34 km s$^{-1}$ at 
145~$\mu$m and 158~$\mu$m respectively. For the 63.2~$\mu$m line non-prime observations
the resolution is $\sim$ 52 km~s$^{-1}$. The instrumental
profiles are represented by Airy profiles whose parameters are
resolution dependent. 

Figure~\ref{f2} presents both the \OI 63.2~$\mu$m and 145.5~$\mu$m lines
together with the \CII 157.7~$\mu$m line toward Sgr~B2 as observed 
with the ISO-LWS Fabry-P\'erot. While the \OI 145.5~$\mu$m line is seen 
in emission, the \CII 158.7~$\mu$m line presents a deep absorption in its 
blue wing whereas the \OI 63.2~$\mu$m is totally absorbed over a large range 
of velocities.

\subsection{IRAM 30-meter observations of $^{13}$CO and C$^{18}$O}

\begin{figure}
\plotone{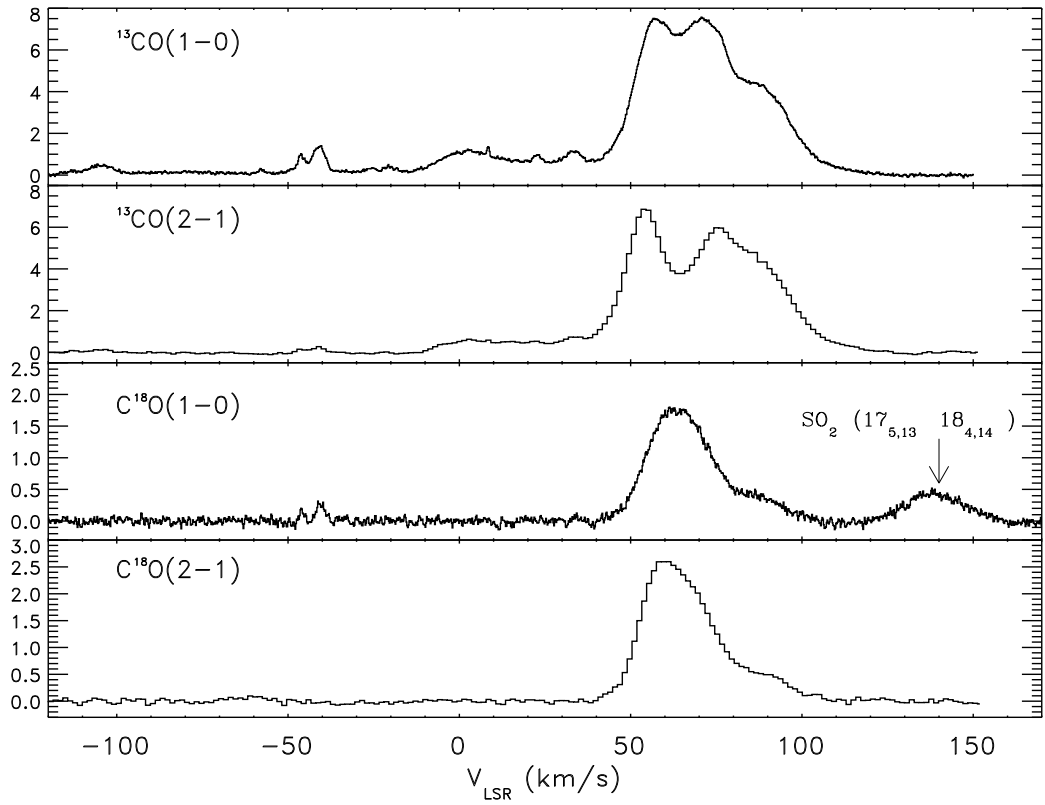}
\caption{$^{13}$CO and C$^{18}$O (1 - 0) and (2 - 1) line spectra degraded to the 80$^{\prime\prime}$
ISO-LWS beam (see text). The y-axis is in main beam temperature.\label{f3}}
\end{figure}

\placefigure{f3}

Molecular observations of Sgr B2 have been carried out in the 
$^{13}$CO (1 - 0) and (2 - 1) and C$^{18}$O (1 - 0) and (2 - 1) 
lines in June 1997 at the IRAM-30 meters telescope (Pico Veleta, Spain). 
The Sgr B2 complex was mapped over the 80$^{\prime\prime}$ LWS beam area 
with a beam size of 21$^{\prime\prime}$ at 2.6 mm and 11$^{\prime\prime}$ at 1.3 mm. 

The spectra show a bright emission profile at the velocity of the HII region 
around 60 km~s$^{-1}$. Many molecular components appear at velocities between 
$-$120 km~s$^{-1}$ and 35 km~s$^{-1}$ in absorption in the central part of the map over a 
30$^{\prime\prime}$ $\times$ 30$^{\prime\prime}$ area, and in emission around. 
The clouds mapped in the LWS beam seem to be homogeneous and uniform as the emission 
components also have their counterparts in the absorption components. We averaged 
the spectra surrounding the 30$^{\prime\prime}$ $\times$ 30$^{\prime\prime}$ area 
and replaced the central spectra by this average. We then convolved the resulting spectra 
with a gaussian weighting function to degrade the spatial resolution
to 80$^{\prime\prime}$ (ISO/LWS beam). 
The constructed spectra are shown in Figure \ref{f3}. 
Up to ten velocity components are detected and their 
parameters, as derived from Gaussian fits, 
are listed in Table \ref{tab1}. 
The upper limits were derived following the relation:

\begin{equation}
3~\sigma~(K) = 3 \times RMS~(K) \times \sqrt{2 \times binsize~(km~s^{-1})/\Delta v~(km~s^{-1})}
\end{equation}

assuming a gaussian profile, where $\Delta v$ is the line width of the
$^{13}$CO or C$^{18}$O (1 - 0) 
transition, RMS and $binsize$ are parameters of the observed spectrum. 
The emission component at $\sim$ 130 km~s$^{-1}$ represents the 
17$_{5,13}$ $\rightarrow$ 18$_{4,14}$ transition of the SO$_2$ molecule.

\placetable{tab1}

One molecular component at $-$76.0 km~s$^{-1}$ only appears in absorption in a 
30$^{\prime\prime}$ $\times$ 30$^{\prime\prime}$ area centered in the $^{13}$CO map 
at the 
coordinates of Sgr B2 (M), which means that the molecular cloud is not extended in 
the beam and/or has a low $^{13}$CO column density. 
This component, not listed in Table 1, 
is averaged over the central positions. It can be fitted by a gaussian with a 
1.4 km~s$^{-1}$ line width and a line to continuum ratio of 0.6 for the $^{13}$CO 
(1 - 0) transition.

\section{Modelling the C$^+$ and O$^0$ Absorption Features}

\subsection{The clouds along the line of sight to Sgr~B2}

\begin{figure}
\epsscale{0.7}
\plotone{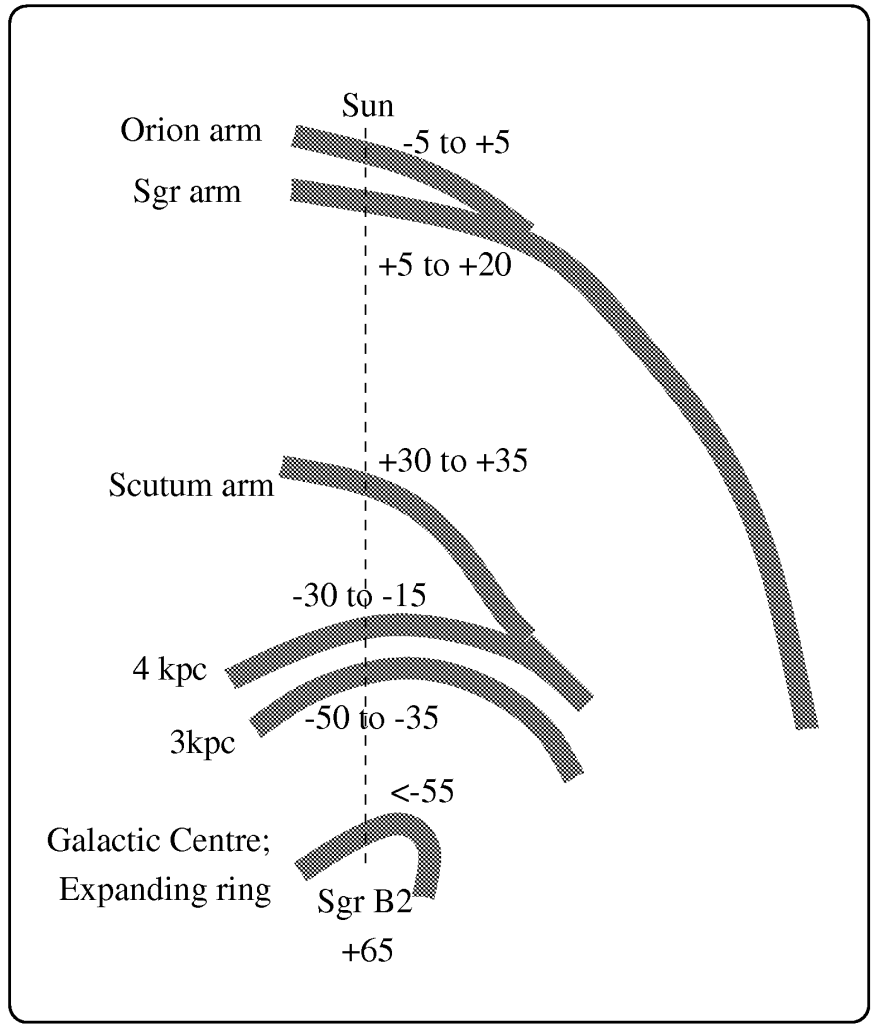}
\caption{Schematic diagram (not to scale) of part of the Galactic Plane, 
showing the position of the background sources, and proposed locations of 
the foreground clouds, with their associated LSR velocities, in 
$\rm km \, s^{-1}$ (adapted from Greaves \& Williams 1994).\label{f4}}
\end{figure}

Individual absorbing clouds can be distinguished in the line of sight only if 
they have a unique radial velocity. For lines of sight away from the galactic 
center the distances to absorbing clouds at different central velocities 
should be estimated from the galactic rotation curve. However, material 
in the Galaxy rotates in orbits deviating from circular ones. 
Figure \ref{f4} presents a sketch of the proposed location of background 
sources and molecular clouds from Greaves \& Williams (1994).

\placefigure{f4}

Due to the large number of velocity components toward Sgr~B2, the determination of the 
location remains difficult and we will adopt, in the following, the Greaves \& 
Williams schematic diagram, except for the component at a velocity around 0~km~s$^{-1}$.
Under circular galactic orbits, any rotating clouds that cross the line of sight could 
contribute to the $\sim$ 0 km~s$^{-1}$ absorption and would be mixed 
with the absorption due to material within a few kpc from the Sun.

The clouds along the line of sight of Sgr~B2 are expected to be illuminated by 
UV photons from the mean interstellar radiation field incident on their external 
layers ($G_0$ = 1 - 10 in units of the standard interstellar flux of 
1.6 $\times$ 10$^{-3}$ erg~s$^{-1}$~ cm$^{-2}$, Habing 1968). 
Following the traditional schematic geometry of the 
photo-dissociation regions (Hollenbach \& Tielens 1999), three different 
major layers can be distinguished in the study of these clouds from 
their surface to their core. Ionized carbon, atomic hydrogen and atomic oxygen coexist 
in the warm, mostly neutral, layer that lies at the surface on which the UV 
impinges. In the next (deeper) layer, where molecular hydrogen is able to resist 
photo-dissociation, ionized carbon and atomic oxygen still coexist. 
The third layer, which constitutes the cold self-shielded molecular core  
of the cloud, contains molecular hydrogen, carbon monoxide and atomic oxygen.

In the present study the \OI absorption at 63.2~$\mu$m is separated into 
two parts for each cloud: the absorption due to atomic oxygen present in the first 
two layers where C$^+$ and O are coexistent (hereafter ''external cloud layers'') 
and that due to atomic oxygen in the cold molecular cores where O and $^{13}$CO 
are coexistent (hereafter "internal molecular cores").  
 
\subsection{The external cloud layers}

\subsubsection{\HI observations}

The diffuse atomic components are parametrized using observations of
the \HI 21~cm line seen in absorption toward Sgr~B2~(M). These 
observations were performed by Garwood \& Dickey (1989) using the VLA 
with a spectral resolution of $\sim$ 5 km~s$^{-1}$ after Hanning
smoothing and a spatial resolution of $\approx 5\arcsec$. We re-fitted 
the absorption components in this spectrum as we found some errors in
the fit parameters quoted by Garwood \& Dickey (1989). 
All absorption features in the observed spectrum were fitted by
Gaussians whose parameters are the central velocities 
($V_{LSR}$), the line widths ($\Delta$v(\HI)) and the \HI optical
depths ($\tau$(\HI)). These parameters are listed in the three first
columns of Table~\ref{tab2} for velocities between $-$110 and +10
km~s$^{-1}$. Higher velocities, corresponding to diffuse \HI clouds
associated with the Sgr~B2 main complex, are not listed as our present 
study concerns only clouds along the line of sight to this complex.  

\subsubsection{The \CII 157.7~$\mu$m line}

\begin{figure}
\plotone{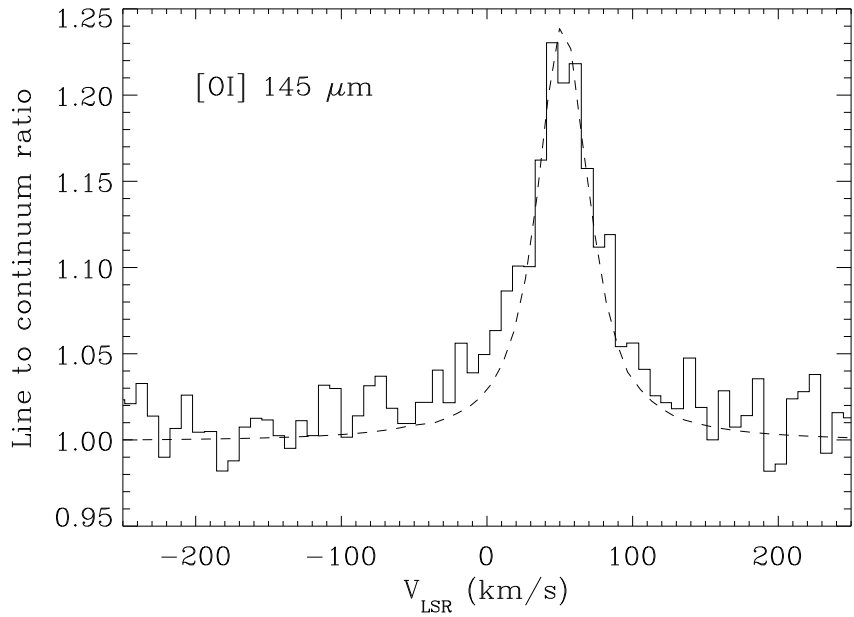}
\caption{Fit of the \OI 145.5~$\mu$m line with a Gaussian of FWHM 
40 km~s$^{-1}$ convolved with the LWS/FP instrumental profile.\label{f5}}
\end{figure}

At  velocities below $\rm 30 \, km \, s^{-1}$, the \CII 157.7~$\mu$m
line is seen in absorption toward Sgr~B2 and in emission at larger
velocities (Fig.~\ref{f2}). 
Before performing any modelling of the absorption feature, an estimate of the 
C$^+$ emission component is required. The physical conditions prevailing 
in the external cloud layers, where C$^+$ is the dominant part of the carbon, 
are assumed to be also valid for atomic oxygen where the \OI 145.5~$\mu$m line 
is emitted, due to the higher gas temperature found in these layers 
(Tielens \& Hollenbach 1985). The \OI 145.5~$\mu$m line (Fig.~\ref{f2}) 
is well fitted with a Gaussian centered around $\rm 60 \, km \, s^{-1}$, 
with an FWHM of $\sim 40 \, \rm km \, s^{-1}$, convolved with the instrumental 
profile. The fit compared to the observations is shown in Fig.~\ref{f5}. 
We will model the C$^+$ emission component by adopting similar parameters
as for the \OI 145.5~$\mu$m line. 

\placefigure{f5}

The C$^+$ absorption features are fitted by Gaussians whose centers 
and line widths are deduced from the \HI 21~cm line parameters. This 
implies that the physical conditions of the layer where hydrogen is 
mainly atomic are also valid for the second layer where H$_2$ is able 
to resist against photo-dissociation. 

The absorption function used for the computations is defined by:

\begin{equation}
I = I_c \times exp{(-\tau(\lambda))},
\end{equation}
\noindent
where I$_c$ is the continuum flux and the optical depth function, $\tau(\lambda)$, is 
defined by:

\begin{equation}
\tau(\lambda) = \tau_0 \times exp\left(-\frac{(\lambda -\lambda_0)^2}{2\sigma^2}\right),
\end{equation}
\noindent
where $\tau_0$ is the optical depth at line center, $\lambda_0$ is the wavelength 
at line center and $\sigma$ is proportional to the width of the absorption lines.

\begin{figure}
\plotone{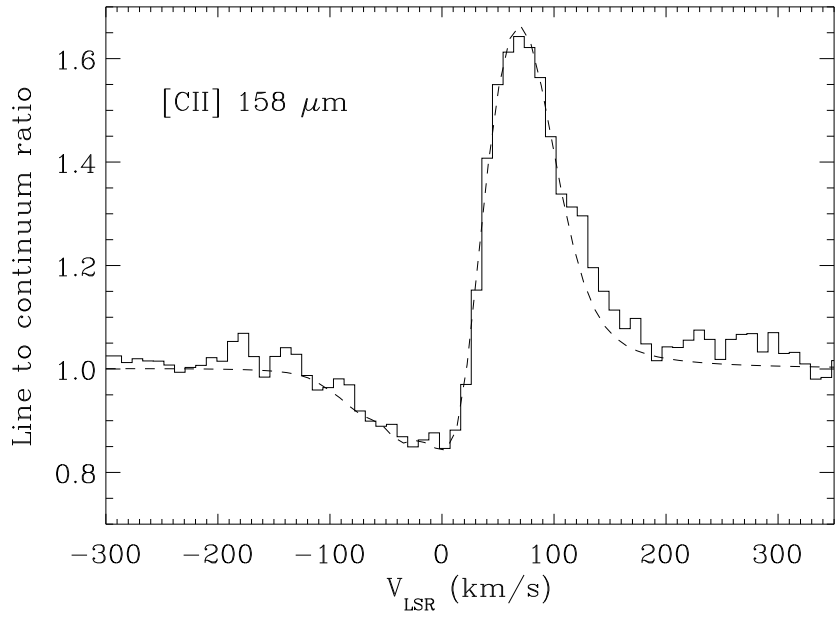}
\caption{The \CII 157.7~$\mu$m line toward Sgr~B2 compared to a model
profile after convolution with the instrumental profile (dashed line). 
See text for details.\label{f6}}
\end{figure}

The optical depth of each C$^+$ absorption feature is adjusted in 
order to reproduce the observed spectrum when combined with the emission 
at 60~km~s$^{-1}$. The final fit of the 157.7~$\mu$m line is shown in 
Fig.~\ref{f6} after convolution of the spectrum with the instrumental 
profile of the LWS-FP at 158~$\mu$m. Note that the fit at velocities above
$\approx  100 \, \rm km \, s^{-1}$ is not perfect, which is due to the lack 
of information at these velocities (see, e.g.,  Garwood and Dickey 1989). 
Another origin of this discrepancy could be due to the transient
effects of the LWS detectors (Caux et al. 2002), not corrected in this
data set.
However, this has no implication for the following results since we are 
concerned with clouds at velocities lower than 10 km~s$^{-1}$. 

\placefigure{f6}

The \CII 157.7 $\mu$m optical depth can be linked to the line width
and to the C$^+$ column density, for a given density and temperature
(see, e.g.,  Crawford et al. 1985). 
For densities lower that 3.3 $\times$ 10$^3$ cm$^{-3}$, a reasonable
estimate of the C$^+$ column density for all temperatures is:

\begin{equation}
N(C^+) \sim 1.3 \times 10^{17} \times \tau_{C^+} \times \Delta v(C^+) \,\, \rm (cm^{-2}),
\end{equation}
\noindent
where the line width $\Delta$v is in km~s$^{-1}$. The computed C$^+$ optical depths and 
column densities are listed in Table~\ref{tab2} (columns 4 and 5, respectively) for 
each component seen in the \HI observations.

\placetable{tab2}

\subsubsection{The \OI 63.2~$\mu$m line}

In the external layers, (as defined above), the major oxygen and carbon
bearing species, in the gas phase, are 
O$^0$ and C$^+$. The variation of their abundances is poorly known at 
galactocentric distances lower than 5 kpc. 
The standard cosmic ratio O/C of 2.3 is used to compute the column 
density of atomic oxygen in the external layers of the clouds (Table~2, column 6). 
This standard ratio is obtained using the cosmic abundance compared to hydrogen 
in the gas phase of 1.4 $\times$ 10$^{-4}$ for carbon (Cardelli et al. 1996) and 
3.2 $\times$ 10$^{-4}$ for oxygen (Meyer, Jura \& Cardelli 1998). The total 
column density of atomic oxygen in the external layers of clouds at $V_{LSR}$ 
lower than 10~km~s$^{-1}$ is estimated to be 8.5 $\times$ 10$^{18}$ cm$^{-2}$.

Since the density in the external layers is believed to be low 
(generally less than 10$^3$~cm$^{-3}$), we can safely assume that, 
in the absorbing region, the majority of oxygen atoms are in the ground 
state. Therefore the column density of atomic oxygen is directly proportional 
to the optical depth (Spitzer 1978) as follows:

\begin{equation}
N(O^0) = \frac{g_l}{g_u}\frac{8\pi}{\lambda^3 A_{ul}}\tau_0 \sqrt{\pi}\frac{FWHM}
{2\sqrt{ln 2}}
\end{equation}
\begin{equation}
\hspace{1cm}= 2.1 \times 10^{17} \times \tau_0 \times \Delta v(O^0) \,\, \rm cm^{-2},
\end{equation}
\noindent
where $g_i$ is the statistical weight of level $i$, $A_{ul}$ = 8.46 $\times$ 10$^{-5}$ s$^{-1}$ 
(Baluja \& Zeippen 1988) is the Einstein coefficient, $\lambda$ = 63.184 $\mu$m, $\tau_0$ 
is the optical depth at line center and $\Delta$v(O$^0$) is the full width at half maximum 
in km~s$^{-1}$ of the absorption line. 

\begin{figure}
\plotone{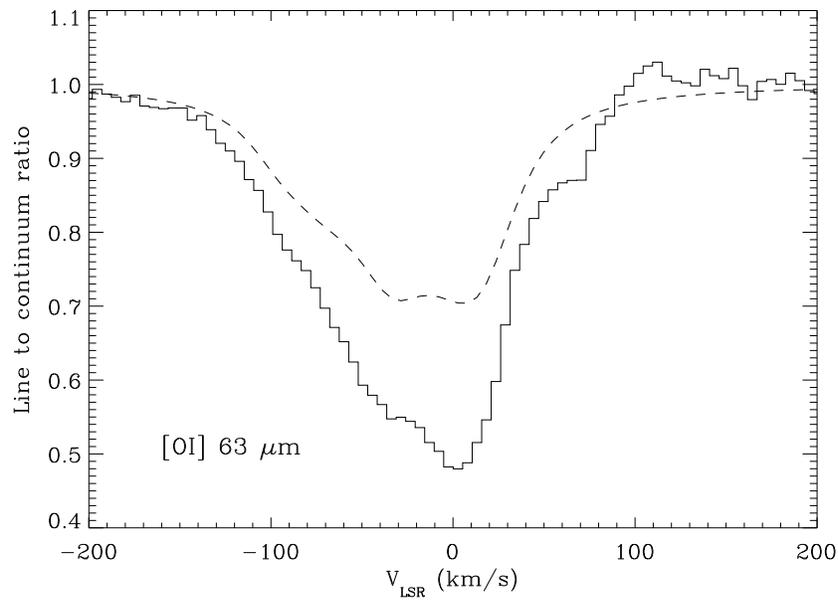}
\caption{The \OI 63.2~$\mu$m line toward Sgr~B2 compared to the predicted
absorption profile for the external layers of foreground clouds ($V_{LSR}$ $<$ 10~km~s$^{-1}$)
shown as  dashed line. The velocity is calculated in the Local Standard of Rest of the 
63.184~$\mu$m line. The lack of absorption around 60~km~s$^{-1}$ is due to the fact 
that the emission component of the source itself is not included in the computations
(see text).\label{f7}}
\end{figure}

In the external part of the clouds, the absorption by atomic oxygen, parametrized by 
$\tau$(O), is computed using the measured \HI line width and the derived column 
density of atomic oxygen (see Table~\ref{tab2}). Above 15 km~s$^{-1}$, due to 
the expected presence of an \OI 63.2~$\mu$m emission line around 60~km~s$^{-1}$ 
from the compact regions in the core of the Sgr~B2 complex, the absorption by atomic 
oxygen in foreground clouds cannot be defined. Again, this has no implication on our results 
since we restrict our study to clouds with velocities lower than 10 km~s$^{-1}$. 
The resulting spectrum is then convolved with the instrumental profile of the 
LWS-FP at 63~$\mu$m and compared to the observations (Fig.~\ref{f7}). 

\placefigure{f7}

A clear result deduced from Fig.~\ref{f7} is that an important fraction of the observed 
absorption cannot be accounted for solely by the external layers of the clouds at velocities lower 
than 10~km~s$^{-1}$. Therefore the contribution of the cold molecular cores to the 
atomic oxygen absorption can now be determined. 
 
\subsection{The internal molecular cores}

Molecules that have been detected in foreground gas clouds not associated with Sgr~B2 
include H$_2$CO (Mehringer, Palmer \& Goss 1995), HCO$^+$, HCN (Linke, Stark \& Frerking 1981), 
C$_3$H$_2$ (Matthews \& Irvine 1985), NH$_3$ (H\"uttemeister et al. 1993), CS, 
C$^{34}$S, H$^{13}$CN, H$^{13}$CO$^+$ and SiO (Greaves et al. 1992), CH (Stacey, 
L\"ugten \& Genzel 1987), H$_2$$^{16}$O and H$_2$$^{18}$O (Neufeld et al. 2000). 
HCN (3 - 2), CS (2 - 1) and (3 - 2) absorption 
lines have been detected toward Sgr~B2 indicating a cloud averaged n$_{H_2}$ density 
close to 200~cm$^{-3}$ and a kinetic temperature between 10 and 20~K (Greaves 1995). 

\placetable{tab3}

From our $^{13}$CO and C$^{18}$O observations, we derive the physical parameters  
of eight molecular cloud cores listed in Table~\ref{tab1} with velocities 
lower than 10~km~s$^{-1}$, using a standard LVG model described in Castets et al. 
(1990). The computations were performed in a restricted range of densities and temperatures 
taking into account the previous estimates through molecular observations. We used 
the $^{13}$CO, C$^{18}$O (1 - 0) and (2 - 1) transitions to 
simultaneously compute the temperature, density and $^{13}$CO column density 
(columns 3, 4, and 5 of Table~\ref{tab3}, respectively). We took into account 
the galactocentric gradient of $^{16}$O/$^{18}$O and $^{12}$CO/$^{13}$CO
in the Galaxy as measured by Wilson \& Rood (1994) and Langer et al. (1990, 1993). At a 
galactocentric distance lower or equal than 4~kpc, we adopted the values $^{12}$CO/$^{13}$CO~=~30 
and $^{13}$CO/C$^{18}$O~=~6. The component near 0~km~s$^{-1}$ is not clearly attributed 
to clouds in the Sgr~B2 complex or to a summation of clouds along the line of sight. 
In this case we used typical values for the local interstellar medium,
i.e. $^{12}$CO/$^{13}$CO~=~60 and  $^{13}$CO/C$^{18}$O~=~10. The kinetic temperature and 
density of the $-$76 km~s$^{-1}$ cloud cannot be computed through our LVG model 
with only the absorption component of the (1 - 0) $^{13}$CO transition. 
Nevertheless, one can estimate the $^{13}$CO column density in this cloud under the 
assumption of local thermodynamic equilibrium (LTE). Dickman (1976) has found that 
the column density of $^{13}$CO derived in LTE is accurate to within a factor of 2 
for dark clouds. N($^{13}$CO) is estimated to be 4.8 $\times$ 10$^{14}$ cm$^{-2}$. 
For the other clouds, we estimate that using the LVG model, the uncertainty on the 
$^{13}$CO column density is lower than 20\%. This value was computed taking into account 
the uncertainties on the best fit of the lines. For the $-$76 km~s$^{-1}$ cloud the 
uncertainty on the $^{13}$CO column density is accurate to within a factor of 2. The final 
column densities with their errors are presented in Table~\ref{tab3}.  

Assuming that the atomic oxygen in the cold molecular component is
coexistent everywhere with $^{13}$CO, the line centers and the line
widths should be the same for the two species. 
Combining the absorption due to both parts of the clouds (the external 
layers and the molecular cores), we can now find the best fit to the
observed absorption by varying the O$^0$ optical depth in each molecular
core and convolving with the instrumental profile. The background
Sgr~B2 source is fitted by an emission 
component at 60~km~s$^{-1}$ for the present computations, corresponding to the 
observed position of the $^{13}$CO and C$^{18}$O line emission (see Sect.~2.2). 
The resulting \OI 63.2 $\mu$m optical thickness and the corresponding atomic oxygen 
column densities, computed using Eq.~6, are reported in Table~\ref{tab3} 
in columns 6 and 7. The total atomic oxygen column density in the molecular 
cores along the line of sight to Sgr~B2 is estimated to be  
$\sim$ 2.2 $\times$ 10$^{19}$~cm$^{-2}$ between $-$120~km~s$^{-1}$ 
and +10~km~s$^{-1}$, i.e. about three times more than that derived
for the diffuse external parts of the clouds. Considering the number
of parameters used in the computation of the atomic oxygen column
density, the uncertainty is difficult to determine. Nevertheless, taking 
into account a 30\% variation in the O$^0$ linewidth and the combined 
systematic and random error of the LWS data, we estimate that the result 
cannot change by more than 50\%. 

\begin{figure}
\plotone{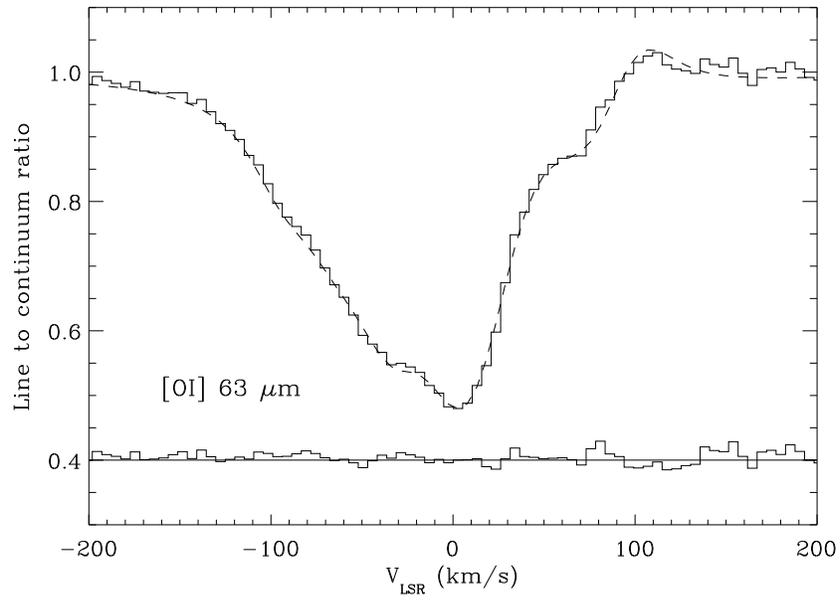}
\caption{Observed \OI 63.2~$\rm \mu m$ absorption profile toward Sgr~B2
together with the best fit (dashed line) including both the external layers 
and the molecular cores contributions. The lower plot shows the
residuals, Y-shifted by +0.4 for clarity.\label{f8}}
\end{figure}

The final result for the combination of the different layers of the 
clouds along the line of sight to Sgr~B2 is shown in Figure~\ref{f8}. 
Additional clouds, responsible for absorption by atomic oxygen at 
velocities larger than 10~km~s$^{-1}$, were introduced only for 
the fit shown in this figure, but their parameters are not reported 
in Table \ref{tab3} as they are clearly dependent on the assumed
\OI 63.2 $\mu$m emission profile and they have no impact on the
results of this present study. 

\placefigure{f8}

The computed O$^0$/$^{13}$CO ratios in the nine molecular components along the line of sight 
of Sgr~B2 are presented in column 8 of Table~\ref{tab3}. In order to compare with 
the canonical value of $\sim 1$  for the O$^0$/$^{12}$CO ratio predicted by chemical models 
(e.g., Lee, Bettens \&  Herbst 1996), we took into account the variation of the isotopic 
ratio $^{12}$CO/$^{13}$CO as function of the galactocentric distance.
The O$^0$/$^{12}$CO ratio (column 9 of Table 3) then is found to range between
3 and 37. The uncertainties are presented in this table, taking into account the combined 
errors on the $^{13}$CO and atomic oxygen column densities.  

\section{Discussion}

A total atomic oxygen column density of 
$\sim$ 3.1 $\times$ 10$^{19}$ cm$^{-2}$ is derived through the  
method developed in this paper, taking into account the diffuse and molecular 
parts of the clouds between $-$120 km~s$^{-1}$ and +6 km~s$^{-1}$. 
This value is consistent with the minimum column density of atomic oxygen computed by 
Baluteau et al. (1997) toward Sgr~B2 of 10$^{19}$ cm$^{-2}$ using the
LWS in its low resolution mode (grating).

As already discussed in Section 3.1, the traditional geometry of 
photo-dissociation regions imply that any cold molecular core should 
be associated with external layers, generally more diffuse and probably 
as halos, that can be 
observed through atomic \HI absorption. We consider in the following 
that this association is effective when the velocity at which 
$^{13}$CO absorption occurs is within the observed line width of the \HI 
absorption feature. Eight such ``clouds'' 
are found along the line of sight to 
Sgr B2, within the velocity range from $-$110 and +10 km~s$^{-1}$, and 
their properties are listed in Table \ref{tab4}. 
The cloud label, the \HI and $^{13}$CO 
associated velocities and the estimated galactocentric distance (see 
Section 3.1) are given in columns 1 to 4.

For each cloud the derived ratio of the neutral oxygen column densities of the 
molecular cores to the whole cloud (external layers + molecular cores) 
is given in column 5. This ratio is indicative of the importance in 
mass of the molecular core(s) within each cloud. Two clouds 
(labelled B and E) are found not to be associated with any significant
molecular core. As indicated in Section 3.3, the molecular cores along the line
of sight to Sgr B2 provide an atomic oxygen column density about
three times that due to the 
diffuse external layers of the clouds.

\subsection{The C$^+$/H ratio in the external layer of the clouds}

The \HI column densities of the diffuse atomic components seen along 
the line of sight to Sgr~B2~(M) can be computed with the standard relation: 

\begin{equation}
N(HI)=1.823 \times 10^{18} \times T_{spin} \times \int \tau(v) dv \hspace{1cm} 
(cm^{-2})
\end{equation}

where (T$_{spin}$) is the \HI spin temperature. 
Through observations of hydrogen absorption features in the direction of 
Sgr B2, Cohen 
(1977) derived the spin temperature of each absorbing cloud.  
However, the radio beam (13$^{\prime}$ by 13$^{\prime}$ angular 
resolution), is much too large compared to the ISO/LWS one, 
and their determination of the spin temperature cannot be used in our study. 
To derive an \HI column density for each cloud along the line of sight to Sgr 
B2, in the range of velocities between $-$110 and +10 km~s$^{-1}$, 
we used a standard value of 150 K for the spin temperature (value well 
within the range derived by Cohen, 1977). The resulting \HI column 
density is given in columns 6 of Table \ref{tab4}. 
The total column density of 
atomic hydrogen in the clouds listed in Table \ref{tab4} 
is about 1.6 $\times$ 10$^{22}$ cm$^{-2}$. 

Assuming that the standard ratio A$_v$/N(H)~=~5.3~$\times$~10$^{-22}$ cm$^{-2}$ 
applies, we calculated that the clouds along the line of sight have 
a diffuse atomic hydrogen surface layer with A$_v$ 
between 0.1 and 4.4 (see column 7 of Table 
\ref{tab4}). It appears that the 
lower the galactocentric distance is, the lower the visual extinction 
of this layer is. This 
could be explained by the higher FUV interstellar radiation field 
encountered at smaller galactocentric distances, leading to a 
sharpening of the atomic \HI layer as indicated by PDR models 
(Hollenbach \& Tielens 1999). However, the identified \HI clouds may be
separated into several smaller clouds if they were viewed at higher
resolution. The computed values of the visual exctinction in each of the 
identified clouds should be taken as an average value and cannot be significantly
indicative of their characteristics.

The computed values for the C$^+$ abundance compared to atomic hydrogen 
(column 8 of Table \ref{tab4}) in the cloud external layers are 
always found larger than the cosmic value in the gas phase of 
1.4 $\times$ 10$^{-4}$ (Cardelli et al. 1996), except in one case 
(cloud E at $-$52 km~s$^{-1}$). This is compatible with the 
picture of 2 external layers: the first one where C$^+$ is coexistent with 
atomic hydrogen and the second one, deeper in the cloud, 
where C$^+$ is coexistent with 
molecular hydrogen. The derived value of C$^+$/H then provides a method to
estimate the mass of the regions filled by C$^+$ and \HI (as the second
layer was not included separately in the calculation). Assuming a C/H
cosmic ratio valid for all the clouds, 
we found that the mass in the C$^+$/\HI and C$^+$/H$_2$ layers should
be approximately the same, which is consistent with what is found for
standard PDRs models (Hollenbach \& Tielens 1999).

\subsection{The C$^0$ layer in the clouds}

\placetable{tab4}

\begin{figure}
\plotone{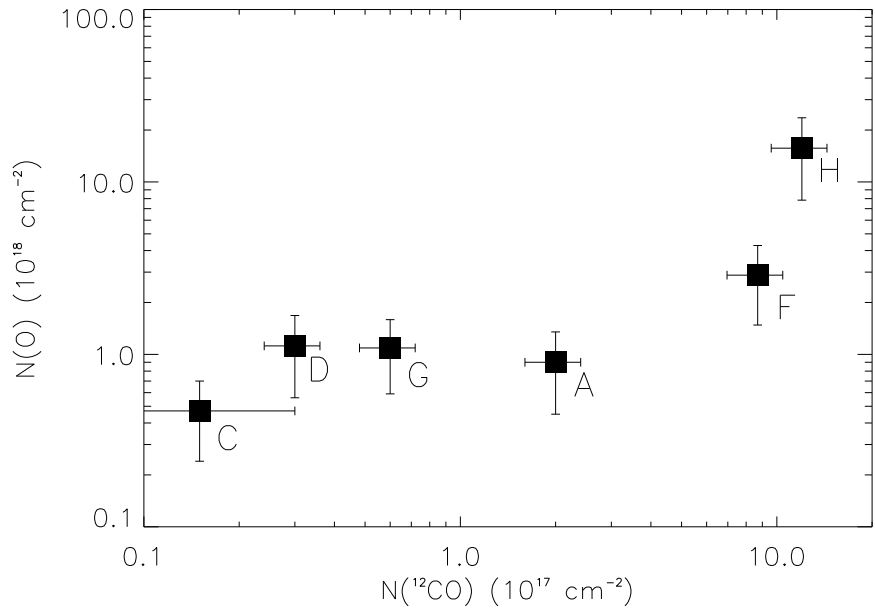}
\caption{The O$^0$ column density against the CO column density for the 
clouds along the line of sight to Sgr B2 associated with molecular 
cores. Cloud labels identify each data point. 
Note that this is a logarithmic scale: the fit was carried out on a linear 
scale. \label{f9}}
\end{figure}

The O$^0$ column densities of the six clouds exhibiting  
molecular core are plotted versus 
their CO column densities in Figure \ref{f9}, with the assumption 
about the galactocentric variation of the $^{13}$C/$^{12}$C ratio as 
given above. 
There is a clear linear correlation between N(O$^0$) and N(CO), except 
for cloud H (at 1.7 km~s$^{-1}$) which is believed to result from a 
blending of clouds with uncertain galactocentric distances.
We performed a linear least-square fit (N(O$^0$)~=~a~+~bN(CO)) taking into 
account that both N(CO) and N(O$^0$) data have errors. This method based on 
the minimization of the $\chi^2$ function leads to the estimation of the a 
and b parameters. The $\chi^2$ function is defined by:

\begin{equation}
\chi^2(a,b) = \sum_{i=1}^{N}\frac{\Delta N(O^0)_i^2}{\sigma_i^2}
\end{equation}  

where $\Delta N(O^0)_i = N(O^0)_i-(bN(CO)_i+a)$ is the difference between the observed 
and fitted value of N(O$^0$), and $\sigma_i^2 = \sigma_{N(O^0)_i}^2 + b^2\sigma_{N(CO)_i}^2$ is 
the weighted sum of the variance in the direction of the smallest $\chi^2$ between each 
data point and the line with slope b. $\sigma_{N(CO)_i}$ and $\sigma_{N(O^0)_i}$ are, 
respectively, the N(CO) and N(O$^0$) standard deviation for the i$^{th}$ point. 
The resulting slope is 2.5 $\pm$ 1.8 and the intersection point 
with the N(O$^0$)-axis gives an excess amount of atomic oxygen with an
average column density of (5.6~$\pm$~2.0)~$\times$~10$^{17}$~cm$^{-2}$. 
This atomic oxygen in excess is interpreted as due to an intermediate layer 
between the C$^+$ external layer and the CO core where the atomic oxygen
coexist with the atomic carbon, as predicted by PDR models.

From theoretical PDR models, a layer of
neutral atomic carbon should exist  near the surfaces of molecular clouds
or clumps within clouds where the UV interstellar radiation field
strikes these surfaces (e. g. Tielens \& Hollenbach 1985) as part
of the C$^+$/C/CO transition region. Observations by 
Plume, Jaffe \& Keene (1994) confirmed that the C$^0$ emission
may arise from PDRs on the surfaces of the molecular clumps distributed
throughout the molecular cloud. As discussed in Tielens \& Hollenbach
(1985) and Hollenbach, Takahashi \& Tielens (1991), the column
density of neutral carbon is insensitive to the strength of the
external FUV field. Only the depth at which the C$^+$/C/CO transition
occurs depends on the FUV field. This strengthens the case for having a
constant atomic carbon column density in all our clouds.\\
Using the cosmic O/C ratio introduced previously with our observations, we deduce 
an average column density for atomic carbon in each cloud along the
line of sight to Sgr B2 of (2.4 $\pm$ 0.9) $\times$ 10$^{17}$ cm$^{-2}$.

This derived C$^0$ column density, found in each cloud associated 
with a molecular core, is well within the range of values derived 
by Usuda et al. ({\it in preparation}) from C$^0$ observations. 
Their first detection of the \CI $^3$P$_1$ - $^3$P$_0$ absorption
lines in the direction of Sgr B2 in the 10.6$^{\prime\prime}$ beam of 
JCMT leads to an estimate of the C$^0$ column density in five cloud 
complexes (clouds labelled here A, C, D, F and G) between 1 and 
13~$\times$~10$^{17}$ cm$^{-2}$. 
 
A summary of previous measurements of the C$^0$/CO ratio in galactic 
interstellar clouds is illustrated in Figure 9 of Usuda et al. ({\it in preparation}) 
where this ratio is plotted against the cloud visual extinction. 
The highest values of this ratio (between 2 and 20) 
are obtained in diffuse 
interstellar clouds (Federman et al. 1980)  
while the lowest values (ranging from 0.03 to 0.3) 
are found in dense photo-dissociation regions (Keene et al. 1985). 
The general slope of the C$^0$/CO vs A$_V$ relation indicates that the 
column density of C$^0$ is relatively constant in clouds with visual 
extinction greater than unity. Our derived values of the C$^0$/CO
ratio are consistent with this general relation; clouds C and D could be 
classified as diffuse clouds (e.g. Federman et al. 1980) where 
C$^0$/CO = [16.2, 8.1], clouds A and G as translucent clouds
(e.g. Stark \& van Dishoeck 1994) where C$^0$/CO = [1.2, 4.0] 
and cloud F as a dense cloud (e.g. Schilke et al. 1995; Maezawa et
al. 1999) where C$^0$/CO = 0.3.

\placefigure{f9}

Values for the C$^+$/C$^0$ ratio in the six clouds associated 
with a molecular core can be derived from this study. While clouds close to the 
galactic center (labelled A, C and D) yield a C$^+$/C$^0$ ratio less than 
0.7, those at galactocentric distance of 3-4 kpc (F and G) have 
ratios close to 2.5. For G$_0$/n $\le$ 3 $\times$ 10$^{-3}$ cm$^3$, the
C$^+$/C/CO transition is drawn to the cloud surface (Kaufman et
al. 1999) which could explain the low C$^+$/C$^0$ ratio ($<$ 1) found near the
galactic center.

\subsection{The O$^0$/CO ratio in the molecular cores}

In the previous Section we provide observational arguments for the presence of 
a neutral carbon layer, at the surface boundary of the molecular cores,
and give estimates of the column density of the associated neutral oxygen. 
Therefore, the actual column density of neutral oxygen in the region where
CO is the major carbon reservoir can be derived, leading to a better
estimate of the true N(O$^0$)/N(CO) ratio in the molecular clouds along
the line of sight to Sgr B2. However, because of the large uncertainties
still present a mean value, derived from the slope of the linear correlation between
N(O$^0$) and N(CO) appears more relevant than individual values. 
Taking into account the errors presented in 
the previous section, it appears that the O$^0$/CO ratio is lower than 4.3 and 
larger than 0.7 with a best fit value of 2.5. It is in good agreement with the 1.8 
value found by Goldsmith (2001: see his Table 3) taking into account a 
molecular depletion in dark cloud cores. Furthermore, this value is within the same 
range of magnitude than the values computed with standard models for dense interstellar 
clouds in steady states (Lee, Bettens \& Herbst 1996). \\

Our derived O$^0$/CO ratio implies that $\sim$ 70\% of gaseous oxygen is in
the atomic form and not locked into CO in the molecular clouds along
the Sgr~B2 line of sight. The upper (respectively lower) limits of this ratio 
corresponds to $\sim$ 80\% (respectively 40\%) of the gaseous oxygen in the 
atomic form. 

The value derived here should be compared to the lower
limit of 15 for the O/CO ratio obtained for molecular clouds along the
W49N HII region line of sight (Vastel et al. 2000). Note that the
\CII 157.7 $\mu$m profile of this source does not present any trace of
either emission or absorption components at the velocities of the
distant clouds. The physical and chemical characteristics of the
clouds in these two lines of sight seem very different and
cannot then be directly connected. Furthermore, the lack of information on the 
intermediate neutral carbon layer in the clouds along the W49N line of sight leads 
to slightly overvalue the O/CO ratio.\\

Considering the calculated uncertainties, we cannot directly determine the 
exact oxygen budget in these molecular clouds. However, only the lower limit on 
the O$^0$/CO ratio is consistent with the predictions of steady states models 
and though we cannot exclude values as low as this (O$^0$/CO $\sim$ 1), the 
probability of larger ratios is much higher.
The oxygen chemistry in dense clouds is still not well known as it is very
difficult to reproduce the O$_2$ and H$_2$O abundances limits measured by
SWAS. Further observations of the O$^0$, O$_2$ and H$_2$O species will 
provide important clues to constrain the chemistry in these cold
molecular clouds.

\subsection{Optical thickness of the C$^+$ line}
 
The \CII fine structure transition at 157.741 $\mu$m is the major 
coolant of the warm interstellar medium. Carbon is the fourth most
abundant element and has a lower ionization potential (11.26 eV) than
hydrogen. Therefore, carbon is in the form of C$^+$ on the surface of
far UV illuminated neutral gas clouds. In addition, the C$^+$ line is
relatively easy to excite, so that the line can efficiently cool the 
warm neutral gas. Depending on the density in this layer, the line may
be self-absorbed. 
The first observational evidence of an absorption feature in the 
C$^+$ spectrum has been carried out by the Kuiper 
Astronomical Observatory toward the W51 HII region (Zmuidzinas,
1987). The absorption
feature in its profile was attributed to a separate ``cooler'' low 
density foreground cloud.
Recently, studies of many bright galaxies showed that the \CII 157.7 
$\mu$m line is deficient compared to the total FIR luminosity. 
Non detection or weak detection have been highlighted toward FIR bright 
galaxies (Stacey et al. 1991; Malhotra et al. 1997; Luhman et al. 1998). 
These observations were carried out at a much lower resolution than
that used in
our study of Sgr B2. Our results indicate that if these galaxies were 
viewed at high spectral resolution, the spectra could reveal both
emission and absorption components present in the C$^+$ profile. This
strengthens the idea that under certain conditions the C$^+$ line
could be optically thick in the direction of large star-forming
complexes or in the nuclei of galaxies.

\section{Conclusion}

Using \CII 157.7 $\mu m$, \OI 63.2 and 145.5 $\mu m$ line observations,
we were able to distinguish between the contributions of the different
layers within the galactic clouds along the line of sight to Sgr~B2. 
We separate the layers of the atomic diffuse surface and of the
molecular core of these clouds. This is a major improvement over the
previous analysis of this line of sight by Lis et al. (2001), which
was based on observations of the \OI 63 $\mu$m line only. We were able 
to associate atomic oxygen with three layers through the clouds 
characterized by different forms of carbon in the gas phase, as
predicted by standard PDR models: i.e. the major form of carbon
changing from C$^+$ to C$^0$ and finally to CO.

From the line shape modelling presented here, a total column density of atomic 
oxygen in the line of sight to Sgr B2 of about 3.1 $\times$ 10$^{19}$ cm$^{-2}$ 
is derived within the clouds with velocities between $-$120 km~s$^{-1}$ 
and +10 km~s$^{-1}$. Less than 30 \% of this total O$^0$ column density is 
found to be due to the external layers of the clouds (where C$^+$ is the 
major form of carbon).

The method used in this study leads to an estimate of the oxygen
content in the intermediate layer where C$^0$ is the dominant form of 
carbon. The atomic carbon column densities derived here for the 
galactic clouds, with a mean value about 2.4 $\times$ 10$^{17}$ cm$^{-2}$, 
are in good agreement with recent observations of the
\CI 492 GHz line. The derived C$^0$/CO ratios are indicative of 
clouds, ranging from diffuse to dense, 
along the line of sight of the Sgr B2 complex which have been 
fragmented and illuminated by the galactic interstellar radiation field. 
C$^+$/C$^0$ ratios close to 2.5 are derived for the clouds at galactocentric 
distances of 3 - 4 kpc, and ratios less than 0.7 for clouds in the 
galactic center region.\\ 
The method used to disentangle the different layers of the clouds in
the line of sight leads to the accurate computation of the O$^0$/CO
ratio ($\sim$ 2.5) in
the internal layers. Therefore, about 70\% of gaseous oxygen is in
the atomic form and not locked into CO in the molecular clouds along
the Sgr~B2 line of sight.   \\
Future instrumentation will enable the present analysis to be improved. The 
Herschel project, with its high spectral resolution (HIFI) capability, will 
allow to better characterize the physical conditions of these clouds along 
the line of sight to Sgr B2 (and maybe on other lines of sight in 
direction of the galactic center as well), in particular from the C$^+$ 
and C$^0$ lines and from high J transitions of CO. SMA and ALMA will provide 
high spatial resolution C$^0$ observations of these clouds necessary
to separate each of them. At high spectral resolution, the
fundamental transitions of atomic oxygen will only be accessible to
the second instrument generation on board the SOFIA observatory.

\acknowledgements
We would like to graciously acknowledge the European Space Agency, the
ISO LWS instrument team and the ISO data reduction software groups. 
We are grateful to Tom Phillips, Jocelyn Keene and Darek Lis for many interesting 
discussions.


\clearpage

\clearpage

\clearpage 

\clearpage

\clearpage

\clearpage

\clearpage

\clearpage

\clearpage

\clearpage

\begin{table}
\caption[]{Observational parameters of the $^{13}$CO and C$^{18}$O (1 - 0) and (2 - 1) 
lines. The 3$\sigma$ upper limit is obtained 
assuming the same line width than that for C$^{18}$O, when detected or 
$^{13}$CO, when C$^{18}$O is not detected. 
The line width corresponds to the $^{13}$CO (1 - 0) line fit.\label{tab1}}
\smallskip
\begin{tabular}{|c|c|c|c|c|c|}
\tableline
v$_{LSR}$($^{13}$CO) & $^{13}$CO (1 - 0)  &$^{13}$CO (2 - 1) 
	& C$^{18}$O (1 - 0)  & C$^{18}$O (2 - 1) & FWHM \\
(km~s$^{-1}$) & T$_{mb}$ (K) & T$_{mb}$ (K) & T$_{mb}$ (K) & T$_{mb}$ (K) & (km~s$^{-1}$)\\
\tableline
$-$113.6 & 0.11 & 0.09    & $<$ 0.04  & $<$ 0.13  & 4.3\\
$-$104.8 & 0.38 & 0.13    & $<$ 0.03  & $<$ 0.10  & 7.5\\
$-$57.8  & 0.18 &$<$ 0.27 & $<$ 0.07  & $<$ 0.21  & 1.6\\
$-$46.2  & 0.78 & 0.12    &    0.17   & $<$ 0.27  & 3.0\\
$-$41.0  & 1.27 & 0.28    &    0.25   & $<$ 0.06  & 4.2\\
$-$25.8  & 0.14 &$<$ 0.23 & $<$ 0.06  & $<$ 0.17  & 2.4\\
$-$20.6  & 0.19 & 0.10    & $<$ 0.06  & $<$ 0.17  & 2.7\\
1.7    & 0.73 & 0.25    & $<$ 0.03  & $<$ 0.07  & 15.0\\
22.7   & 0.30 & 0.10    & $<$ 0.06  & $<$ 0.06  & 2.1\\
33.3   & 0.51 & 0.23    & $<$ 0.04  & $<$ 0.13  & 4.5\\
\tableline
\end{tabular}
\end{table}

\clearpage

\begin{table}
\caption[]{The derived parameters for the external layers of the clouds 
associated with \HI absorption (Garwood and Dickey 1989) along the line of 
sight to Sgr~B2 in the range of velocities between $-110$ and +10~km~s$^{-1}$.\label{tab2}}
\smallskip
\begin{tabular}{|c|c|c|c|c|c|}
\tableline
V$_{LSR}$    & $\Delta$v(\HI) & $\tau$(\HI) & $\tau$(C$^+$) & N(C$^+$)  & N(O$^0$) \\
(km~s$^{-1}$)& (km~s$^{-1}$) & & & (10$^{17}$ cm$^{-2}$) & (10$^{17}$ cm$^{-2}$) \\
\tableline
$-108$   & 7.0  & 0.14& 0.10 & 1.3 & 3.0  \\
$-92$    & 14.0 & 0.13& 0.08 & 1.5 & 3.5  \\
$-77$    & 14.0 & 0.19& 0.10 & 1.8 & 4.1  \\
$-60.5$  & 7.0  & 0.3 & 0.10 & 0.9 & 2.1  \\
$-51.9$  & 8.0  & 0.55& 0.10 & 1.0 & 2.3  \\ 
$-44$    & 8.0  & 1.1 & 0.60 & 6.2 & 14.3 \\
$-21.5$  & 15.0 & 0.5 & 0.30 & 5.9 & 13.6 \\
$-3.5$   & 11.5 & 1.4 & 0.50 & 7.5 & 17.3 \\
$5.5$    & 12.0 & 1.2 & 0.70 & 10.9& 25.1 \\
\tableline
\end{tabular}
\end{table}

\clearpage

\begin{table}
\caption[]{The derived parameters for the  
nine molecular cloud cores along the line of sight to Sgr~B2 
in the range of velocities between $-110$ and +10 km~s$^{-1}$.\label{tab3}}
\smallskip
\begin{tabular}{|c|c|c|c|c|c|c|c|c|}
\tableline
Position & FWHM  & T  &  n$_{H_2}$  & N($^{13}$CO) & $\tau_0$(O$^0$)  & N(O$^0$) & O$^0$/$^{13}$CO & O$^0$/$^{12}$CO\\
(km~s$^{-1}$) & (km~s$^{-1}$) & (K) & (cm$^{-3}$) & (10$^{15}$ cm$^{-2}$) &   &  (10$^{18}$ cm$^{-2}$) & &\\
\tableline
$-113.6$ & 4.32  & 10  &  200        & 1.7 $\pm$ 0.3 &  0.3  & 0.27 $\pm$ 0.13 &  159 $_{-89}^{+127}$  & 5 $_{-3}^{+4}$\\
$-104.8$ & 7.47  & 10  &  600        & 5.0 $\pm$ 1.0 &  0.7  & 0.63 $\pm$ 0.31 &  126 $_{-73}^{+109}$  & 4 $_{-2}^{+4}$\\
$-76.0$  & 1.40  &     &             & 0.5 $\pm_{0.25}^{0.5}$ &  1.7  & 0.47 $\pm$ 0.23 &  940 $_{-700}^{+1860}$  & 31 $_{-23}^{+62}$\\
$-57.8$  & 1.60  & 10  &  200        & 1.0 $\pm$ 0.2 &  3.3  & 1.12 $\pm$ 0.56 &  1120 $_{-653}^{+980}$ & 37 $_{-22}^{+33}$\\
$-46.2$  & 2.96  & 10  &  300        & 6.0 $\pm$ 1.2 &  1.8  & 1.12 $\pm$ 0.56 &  187 $_{-114}^{+163}$  & 6 $_{-4}^{+6}$\\
$-41.0$  & 4.20  & 10  &  200        & 23.0$\pm$ 4.6 &  2.0  & 1.76 $\pm$ 0.88 &  77 $_{-45}^{+67}$   & 3 $_{-2}^{+4}$\\
$-25.8$  & 2.44  & 10  &  200        & 1.0 $\pm$ 0.2 &  1.0  & 0.51 $\pm$ 0.25 &  510 $_{-293}^{+440}$  & 17 $_{-10}^{+15}$\\
$-20.6$  & 2.74  & 10  &  200        & 1.0 $\pm$ 0.2 &  1.0  & 0.58 $\pm$ 0.29 &  580 $_{-338}^{+508}$  & 19 $_{-11}^{+17}$\\
$1.7$    & 14.97 & 15  &  300        & 20.0$\pm$ 4.0 &  5.0  & 15.72 $\pm$ 7.86 &  786 $_{-459}^{+688}$  & 13 $_{-12}^{+8}$\\
\tableline
\end{tabular}
\end{table}

\clearpage

\begin{table}
\small
\caption{Compared properties of the  
eight clouds found along the line of sight to Sgr B2 
in the range of velocities between $-$110 and +10 km~s$^{-1}$. 
A correlation between N(O$^0$) and N($^{12}$CO) is established in
figure \ref{f9} (see text).\label{tab4}}
\smallskip
\begin{tabular}{|c|c|c|c|c|c|c|c|c|}
\tableline
Cloud & external layers & molecular cores & R$_G$ & N(O$^0$) ratio & N(\HI) &
A$_v$ & C$^+$/H & O$^0$/$^{12}$CO\\
label & V$_{LSR}$ (km~s$^{-1}$) & V$_{LSR}$ (km~s$^{-1}$) & (kpc) &
core/total  & (10$^{20}$ cm$^{-2}$) &  &
(10$^{-4}$)  & \\
\tableline\tableline
A & $-$108 & $-$113.6/$-$104.8  & $<$ 1   & 75\%  & 2.7  & 0.1  & 4.8   & 5 $_{-3}^{+4}$ \\
B & $-$92  &                & $<$ 1   & 0\%   & 5.0  & 0.3  & 3.0   &   \\
C & $-$77  &  $-$76           & $<$ 1   & 53\%  & 7.3  & 0.4  & 2.5   & 31 $_{-23}^{+62}$ \\
D & $-$60.5 & $-$57.8         & $<$ 1   & 84\%  & 5.7  & 0.3  & 1.6   & 37 $_{-22}^{+33}$ \\
E & $-$51.9 &               &  3      & 0\%   & 12.0 & 0.6  & 0.8   &   \\
F & $-$44 & $-$46.2/$-$41.0     &  3      & 67\%  & 24.1 & 1.3  & 2.6   & 3 $_{-2}^{+3}$ \\
G & $-$21.5 & $-$25.8/$-$20.5   &  4      & 44\%  & 20.5 & 1.1  & 2.9   & 18 $_{-10}^{+15}$ \\
H & $-$3.5/5.5 & 1.7        &  ?      & 79\%  & 83.4 & 4.4  & 1.7   & 13 $_{-8}^{+12}$\\
\tableline
\end{tabular}
\end{table}

\clearpage

\end{document}